# Distraction by auditory novelty during reading: Evidence for disruption in saccade planning, but not saccade execution


Martin R. Vasilev[1] [*]

Fabrice B. R. Parmentier[2,3,4]

Julie A. Kirkby[1]

[1]Bournemouth University, Department of Psychology, United Kingdom

[2]Department of Psychology and Research Institute for Health Sciences (iUNICS), University of the Balearic Islands, Palma, Balearic Islands, Spain

[3]Balearic Islands Health Research Institute (IdISBa), Palma, Balearic Islands, Spain

[4]School of Psychology, University of Western Australia, Perth, Western Australia, Australia

Corresponding author[*] at:

Department of Psychology, Bournemouth University

Poole House, Talbot Campus, Fern Barrow

Poole, Dorset, BH12 5BB, United Kingdom

Phone: +44 7835202606

Email: mvasilev@bournemouth.ac.uk


Word count: 7629 words



Abstract

Novel or unexpected sounds that deviate from an otherwise repetitive sequence of the same sound cause behavioural distraction. Recent work has suggested that distraction also occurs during reading as fixation durations increased when a deviant sound was presented at the fixation onset of words. The present study tested the hypothesis that this increase in fixation durations occurs due to saccadic inhibition. This was done by manipulating the temporal onset of sounds relative to the fixation onset of words in the text. If novel sounds cause saccadic inhibition, they should be more distracting when presented during the second half of fixations when saccade programming usually takes place. Participants read single sentences and heard a 120 ms sound when they fixated five target words in the sentence. On most occasions ($p= 0.9$), the same sine wave tone was presented ("standard"), while on the remaining occasions ($p= 0.1$) a new sound was presented ("novel"). Critically, sounds were played either during the first half of the fixation (0 ms delay) or during the second half of the fixation (120 ms delay). Consistent with the saccadic inhibition hypothesis, novel sounds led to longer fixation durations in the 120 ms compared to the 0 ms delay condition. However, novel sounds did not generally influence the execution of the subsequent saccade. These results suggest that unexpected sounds have a rapid influence on saccade planning, but not saccade execution.

Word count: 231 words



Novel or unexpected sounds that occur in an otherwise repetitive sequence of the same sound cause cognitive distraction that is observed in both neurophysiological and behavioural responses (Berti, 2012; Escera, Alho, Schröger, & Winkler, 2000; Horváth, Winkler, & Bendixen, 2008; Parmentier, 2014; Parmentier, Vasilev, & Andrés, 2019; Schröger, 1996). The detrimental effect of unexpected sounds on behavioural performance has been reported using visual, auditory and crossmodal categorization tasks, two-alternative forced-choice tasks, Go/ NoGo, visual matching, and serial recall tasks (Bendixen et al., 2010; Berti & Schröger, 2001; Hughes, Vachon, & Jones, 2005; Li, Parmentier, & Zhang, 2013; Ljungberg & Parmentier, 2012; Ljungberg, Parmentier, Leiva, & Vega, 2012; Pacheco-Unguetti & Parmentier, 2016; Parmentier, 2016; Röer, Bell, Körner, & Buchner, 2018; Röer, Bell, Marsh, & Buchner, 2015). Interestingly, recent research has suggested that unexpected sounds may also lead to longer fixation durations during everyday tasks such as reading and scene viewing (Graupner, Velichkovsky, Pannasch, & Marx, 2007; Vasilev, Parmentier, Angele, & Kirkby, 2019). In the present study, we tested the hypothesis that the increase in fixation durations during reading is due to inhibition of saccade planning. Additionally, we examined whether novel sounds affect the execution of reading saccades.

## Distraction by Auditory Novelty

Novel sounds are distracting not because of their low frequency of occurrence, but because they violate sensory predictions (Bubic, von Cramon, Jacobsen, Schröger, & Schubotz, 2009; Parmentier, Elsley, Andrés, & Barceló, 2011; Schröger, Bendixen, Trujillo-Barreto, & Roeber, 2007), even when the stimuli and responses in the primary task are predictable (Parmentier & Gallego, 2020). Novelty distraction is associated with specific electrophysiological responses: 1) a mismatch negativity (MMN) component, reflecting an early detection of auditory change in the brain (Näätänen, Gaillard, & Mäntysalo, 1978; Näätänen, Paavilainen, Rinne, & Alho, 2007); 2)



a P3a component, reflecting the involuntary orientation of attention away from the main task (Berti, 2012; Escera et al., 2000; Schröger & Wolff, 1998); and 3) a re-orientation negativity (RON) component, interpreted as the re-orientation of attention to the main task (Berti & Schröger, 2001; Horváth et al., 2008; Schröger, Giard, & Wolff, 2000). Hence, novelty distraction is typically viewed as the outcome of an involuntary orienting response (Sokolov, 1963) towards the novel sound. The shift of attention to and away from the novel sound, as well as the processing aftermath of its involuntary semantic analysis, have been argued to be important determinants of the reduction in main task performance (Parmentier, Elford, Escera, Andrés, & Miguel, 2008; Parmentier, Turner, & Perez, 2014; Schröger, 1996).

Interestingly, recent evidence has suggested that novel sounds may also lead to general motor inhibition (Dutra, Waller, & Wessel, 2018; Wessel, 2017; Wessel & Aron, 2013, 2017). For example, Wessel and Aron (2013) found that novel sounds led to a reduction in cortico-spinal excitability some 150 ms after their presentation, which was interpreted as evidence for global inhibition of the motor system. This reduction in cortico-spinal excitability appears to be directly related to motor planning as it was positively correlated with action stopping in a Go/NoGo task (Dutra et al., 2018). These results suggest that unexpected novel sounds may trigger a transient and rapid inhibition of motor responses. Such inhibition may occur earlier in time than the attention orienting response (Wessel & Aron, 2017) and stop ongoing processes to enable a more rapid and effective analysis of unexpected sounds (Wessel, 2017).

While novelty distraction is typically measured from response times using forced-choice categorization tasks, recent evidence has suggested that deviant sounds may also affect oculomotor control (Graupner et al., 2007; Vasilev et al., 2019; Widmann, Engbert, & Schröger, 2014; see also Marois & Vachon, 2018; Wetzel, Buttelmann, Schieler, & Widmann, 2016 for



pupil dilation responses). For example, Graupner et al. (2007) presented visual and auditory distractors at every fifth fixation in a scene viewing task. Participants were presented with 17 standard distractors, 16 standard distractors and one deviant distractor, or no distractors. Graupner et al. (2007) found that the deviant sound led to an increase in fixation durations, which was due to a reduction in the proportion of terminated fixations. This reduction occurred at two distinct time intervals: first at around 90 ms and then at around 150 ms after the sound's onset.

Additionally, Widmann et al. (2014) studied the categorisation of sounds using *microsaccades*, which are miniature eye-movements that occur about 1-2 times per second (Engbert, 2006). Widmann et al. (2014) found a significant difference between standard and target sounds starting at 142-148 ms, and between distractors (intensity/ pitch deviants) and target sounds starting at 148-196 ms after sound onset. After factoring in neural transmission and motor delays, they argued that microsaccades can show the categorisation of target vs non-target sounds some 80-100 ms after sound onset.

Furthermore, Vasilev et al. (2019) presented a 50-ms sound when readers fixated five target words in a sentence. Participants either heard five standard sounds (a sine wave) or four standard and one deviant sound (a burst of white noise). The authors found that the deviant sound led to an increase in fixation durations immediately after its presentation. A time-course analysis revealed that the deviant sound began to affect fixation durations some 180 ms following the sound's onset. Because the increase in fixation durations originated relatively late in the fixation duration distribution and did not appear to be related to the lexical processing of words in the sentence, Vasilev et al. (2019) hypothesised that this delay stemmed from the disrupted programming of the next saccade. In summary, these results suggest that unexpected sounds can affect oculomotor control during everyday tasks such as scene viewing and reading.



**Eye-movement Control during Reading**

During reading, the eyes alternate between short periods of relative stability (i.e., *fixations*) and quick, ballistic movements (i.e., *saccades*). While fixations allow readers to uptake high-resolution visual information from the current word, saccades bring their eyes to unexplored parts of the text. There is now a large body of evidence showing that fixation durations during reading are sensitive to different linguistic properties of words, such as their lexical frequency (Inhoff & Rayner, 1986; Rayner & Duffy, 1986; Schilling, Rayner, & Chumbley, 1998) or predictability given the preceding context (Balota, Pollatsek, & Rayner, 1985; Kliegl, Nuthmann, & Engbert, 2006; Rayner, Slattery, Drieghe, & Liversedge, 2011; Rayner & Well, 1996; see Rayner, 1998, 2009 for a review). This suggests that fixation durations are influenced by the cognitive processing of the text.

The average fixation duration during reading is about 240 ms (Reichle & Reingold, 2013). In that time, readers have to extract linguistic information from the fixated word in order to recognise it and then plan a saccade to the next word. There is evidence that word processing starts soon after fixation onset. For example, visual input during the first 60 ms appears to be crucial for word recognition as lexical processing occurs even if the word disappears afterwards (Liversedge et al., 2004; Rayner, Liversedge, White, & Vergilino-Perez, 2003). Additionally, linguistic variables typically have a quick influence on fixation durations. For instance, survival analyses have shown that lexical frequency, predictability, lexical ambiguity, and preview validity start to influence fixation durations between 120-140 ms from fixation onset (Reingold & Sheridan, 2014). Moreover, neurophysiological evidence has shown that lexical processing typically occurs around 127- 172 ms after fixation onset on average (Reichle & Reingold, 2013). However, because some parafoveal pre-processing usually occurs before words are fixated,



lexical processing often starts even earlier than that (Reichle & Reingold, 2013). Therefore, lexical processing of the fixated word typically begins prior to the planning of the next saccade.

This idea is illustrated well in serial-attention models of eye-movement control such as E-Z Reader (Reichle, Pollatsek, Fisher, & Rayner, 1998; Reichle, Warren, & McConnell, 2009). In this model, word recognition starts with a 50-ms visual processing stage that reflects the time needed for visual information from the retina to reach the cortex (Pollatsek, Reichle, & Rayner, 2006; see also Foxe & Simpson, 2002). This is then followed by two attention-dependent lexical processing stages- familiarity check (L1) and lexical access (L2). L1 roughly corresponds to the recognition of the orthographic wordform and is used to estimate the difficulty of accessing the meaning of the word. L2, on the other hand, reflects the actual act of word identification (i.e., recognising and retrieving the word's meaning from memory; Reichle, Rayner, & Pollatsek, 2003). Once L1 is completed, the programming of the saccade to the next word is initiated because completion of the second stage (L2) is imminent. The programming of the next saccade also occurs in two stages: 1) a labile stage (M1) in which the current saccade plan can be cancelled by another programme; and 2) a non-labile stage (M2) in which the current saccade plan can no longer be cancelled. Once lexical processing of the fixated word is completed, attention shifts covertly to the next word in anticipation of the eye-movement.

Parallel-attention models such as SWIFT (Engbert, Longtin, & Kliegl, 2002; Engbert, Nuthmann, Richter, & Kliegl, 2005) differ from E-Z Reader in that attention is allocated to more than one word at a time, but they also share a number of core assumptions. For instance, lexical processing in SWIFT also occurs in two stages, as does the programming of the next saccade. However, contrary to E-Z Reader, the trigger to start planning the next saccade in SWIFT is not determined by the completion of some initial lexical processing, but is instead based on a random



saccadic timer with a pre-defined mean (Engbert et al., 2005). Nevertheless, this timer can be inhibited foveally by the processing difficulty of the fixated word, such that less frequent (i.e., more difficult) words will prolong fixation durations. This effectively delays the onset of the next saccade to allow for lexical processing of less frequent words to occur. Such delay mechanism could only work if some lexical processing occurs early enough to allow the difficulty of the fixated word to be estimated and the programming of the next saccade to be delayed as a consequence. Therefore, in both models, lexical processing of the fixated word must generally start early on in the fixation, whereas saccade planning by necessity occurs towards the end of fixations.

**Neural Control of Saccades**

The execution of saccades occurs immediately after their planning. However, it is currently not known whether saccade execution can be perturbed by novel sounds. Saccades begin with an acceleration phase that lasts until their peak velocity is reached, after which they start to decelerate. The duration and peak velocity of saccades increases non-linearly with greater saccadic amplitude (often called the "main sequence"; Bahill, Clark, & Stark, 1975). Reading saccades usually have an amplitude of ~ 2º (7-9 letters; Rayner, 2009) and last for about 20-40 ms (Bouma & De Voogd, 1974; Pollatsek et al., 2006). Saccade execution is controlled by brainstem burst neurons, which receive their main input from the superior colliculus (SC) (Sparks, 2002; Watanabe & Munoz, 2011). Importantly, saccadic parameters are coupled to the activity of excitatory burst neurons (EBNs) in the brainstem. More specifically, peak saccade velocity is correlated with the maximum firing rate of EBNs, saccade duration is correlated with their burst duration, and the number of spikes in the burst is correlated with saccade amplitude (Fuchs, Kaneko, & Scudder, 1985; Galley, 1989; Sparks, 2002). Therefore, as the execution of



saccades depends on the firing of brainstem neurons, any inhibition of these neurons should also be reflected in slower saccade velocities and longer saccade durations.

**Present Study**

Unexpected sounds lead to an immediate increase in fixation durations during reading (Vasilev, Parmentier, et al., 2019), which may occur due to global transient inhibition of motor responses (Wessel & Aron, 2013, 2017). Therefore, the increase in fixation durations may be due to saccadic inhibition during the planning stages of the next saccade. We will refer to this explanation as the *saccadic inhibition hypothesis* (SIH). This hypothesis is supported by the finding that deviance distraction occurs late in fixation duration distributions and does not appear to be modulated by the lexical frequency of words (Vasilev, Parmentier, et al., 2019). However, this evidence is only indirect in nature. Here, we set out to test the SIH more directly by manipulating the timing of auditory distractors (standard versus novel) relative to the fixation onset of words.

As already mentioned, lexical processing of words starts soon after fixation onset, whereas programming of the next saccade by necessity occurs towards the end of fixations. Therefore, if the SIH is correct, novel sounds should be more distracting when played temporarily closer to the end of fixation (i.e., when saccade programming takes places). To test this hypothesis, we manipulated the temporal onset of sounds relative to the fixation onset of words. A 120 ms sound was played when participants fixated five target words in a sentence. The same sine wave tone ("standard") was played on most occasions (p=.9), while a new environmental sound ("novel") was played on the remaining occasions (p=.1). Critically, and orthogonally to this sound manipulation, the sound's onset either coincided with the fixation's onset (0 ms delay) or followed it by 120 ms (120 ms delay). Because the average reading fixation



is about 240 ms long (Reichle & Reingold, 2013), this timing manipulation ensured that the sound occurred, on average, either during the first half or the second half of the fixation. Therefore, under the SIH, novel sounds should lead to significantly longer fixation durations than standard sounds in the 120 ms delay condition compared to the 0 ms delay condition.

A secondary goal of our study was to examine the unexplored issue of whether novel sounds may affect not only the programming, but also the execution of the next saccade. This is a relevant question given both the temporal overlap between saccade planning and execution, and the anatomical overlap between structures thought to be implicated in the temporary suppression of motor activity and structures controlling the execution of saccades. Wessel and Aron (2013) have argued that unexpected events lead to a global inhibition of the motor system, which may be modulated by the subthalamic nucleus (STN). The STN contains visual-motor neurons (Fawcett, Dostrovsky, Lozano, & Hutchison, 2005; Matsumura, Kojima, Gardiner, & Hikosaka, 1992) and receives direct projections from the cerebral cortex (Ma & Geyer, 2017). Furthermore, the STN may be involved in maintaining a sustained fixation before goal-directed saccades by suppressing neurons in the SC via an indirect route through the substantia nigra pars reticulata (SNr) (Hikosaka, Takikawa, & Kawagoe, 2000). Of interest, the SC also controls the execution of saccades in the brainstem (Sparks, 2002; Watanabe & Munoz, 2011). Therefore, it is possible, but yet unknown, that the motor inhibition thought to follow unexpected sounds may also affect the firing of EBNs, which is related to the peak velocity of saccades (Fuchs et al., 1985). If novel sounds affect the execution of saccades, then we should observe a reduction in their velocity and an increase in their duration. This should particularly be the case in the 120 ms delay condition, where the novel sound is played temporarily closer to the execution of the next saccade.

**Hypotheses**



**H1:** Consistent with previous results (Vasilev, Parmentier, et al., 2019), novel sounds should result in longer fixation durations compared to standard sounds (i.e., main effect of Sound).

**H2**: Consistent with the SIH, novel sounds should be more distracting when played during the second half of fixation when the next saccade is usually being programmed (i.e., Sound x Delay interaction).

**H3.1:** If novel sounds also affect the execution of the next saccade, they should result in a decrease in saccade velocity and an increase in saccade durations. Additionally, these effects should be stronger in the 120 ms compared to the 0 ms delay condition because the former is temporally closer to the execution of the next saccade (**H3.2**).

## Method

### Participants

Sixty-four undergraduate students from Bournemouth University (53 female) participated for course credits. Their average age was 19.3 years (range: 18-31 years; *SD*= 2.12 years)[1]. Participants were native English speakers who reported normal or corrected-to-normal vision, normal hearing, and no prior diagnosis of reading disorders. Participants were naïve as to the purpose of the experiment. The study was approved by the Bournemouth University Research Ethics Committee and all participants provided informed written consent.

### Materials and Design

---

[1] Four more participants were tested but replaced: one due to equipment failure and three due to tracking problems caused by wearing glasses or contact lenses.



The method is illustrated in Figure 1. The reading stimuli consisted of 120 English sentences that were taken from Vasilev et al. (2019). In each sentence, there were five target words on which the sounds were presented using a gaze-contingent manipulation (see Inhoff, Connine, & Radach, 2002). The target words were always the third, fifth, seventh, ninth and eleventh word in the sentence. Each target word was followed by one non-target word on which no sounds were played. This ensured that the sounds were not played too quickly after one another. The target words were 6.75 letters long on average (SD= 1.89 letters; range: 3- 13 letters). Short function words were avoided as targets to increase the likelihood that the target words would be fixated during first-pass reading.

Sixty novel sounds were taken from Andrés, Parmentier, and Escera (2006). The sounds were originally 200 ms long but were compressed to 120 ms (without alteration in pitch) for the purpose of the present experiment. The novel sounds consisted of different environmental sounds (e.g., sound of a drill, telephone ringing, engine, etc.). The standard sound was a 120 ms sinewave tone with a frequency of 400 Hz, with 10 ms fade-in and fade-out ramps. All sounds were monoaural, with a sampling rate of 44100 Hz, a bit depth of 16 bit. A silence condition was not included since (1) we previously showed that the presentation of five gaze-contingent standard sounds does not affect reading behaviour compared to silent reading (Vasilev, Parmentier, et al., 2019); and (2) because it was not relevant to our objective, which was to determine whether the timing of the novel sounds influences oculomotor control.



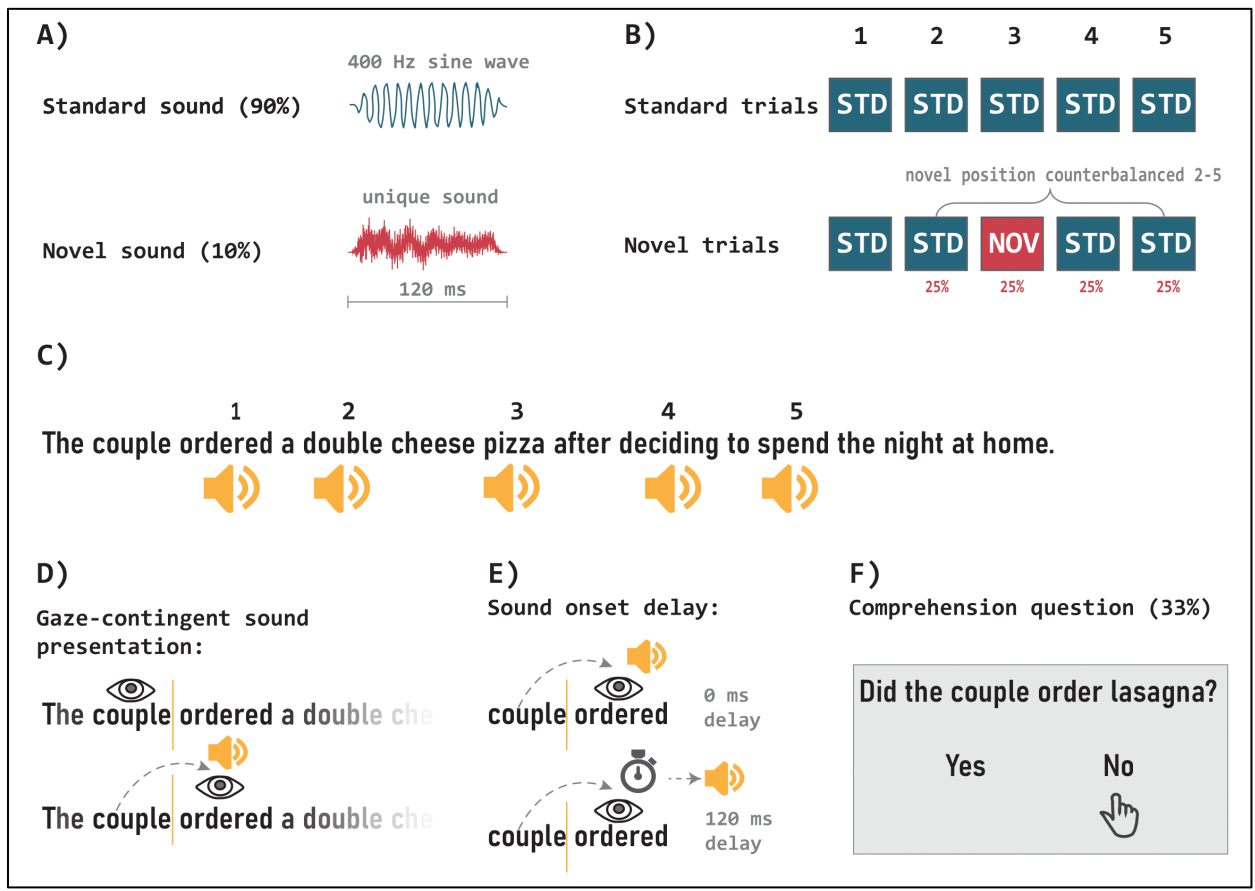

*Figure 1*. Diagrammatic illustration of the task's key characteristics. **A**: Standard and novel sounds used in the experiment. **B**: Presentation of sounds in standard and novel trials. Standard trials contained five standard sounds, while novel sounds contained four standard and one novel sound (played equiprobably at positions 2-5). **C**: Example sentence with the position of words on which the five sounds were played. **D**: Gaze-contingent sound presentation in the experiment. Sounds were played when participants crossed an invisible boundary (Rayner, 1975; denoted by a vertical orange line) located before the five target words. **E**: Sound onset delay manipulation. Sounds were played either immediately (0 ms delay) or with a 120 ms delay after crossing the boundary. The delay was the same for all sounds played in a given sentence. **F**: Example of a comprehension question (presented after 33% of sentences).

The study had a 2 x 2 within-subject design, with *sound type* (standard vs novel) and *sound onset delay* (0 vs 120 ms) as the factors. Each participant saw 30 trials per condition. There



were two types of trials that corresponded to the two sound conditions. In *standard* trials (which represented 50% of all trials), the standard sound was played on each of the five target words in the sentence. In contrast, in *novel* trials (remaining 50% of the trials), one novel and four standard sounds were played. Novel trials always started with a standard sound on Target Word 1 to reactive the representation of the standard sound at the beginning of the trial, as this representation may weaken or become "dormant" due to the pause between trials (e.g., Cowan, Winkler, Teder, & Näätänen, 1993; for a review, see Winkler & Schröger, 2015). The novel sound was then presented on one of the four remaining Target Words (2 through 5) with equal probability across the experiment. Novel sounds made up 10% of all sounds played during the experiment. In each sound condition, half of the sounds were played with no delay (i.e., 0 ms) and the other half were played with 120 ms delay after the target word was fixated. The delay was always the same for the five sounds played in each sentence. The assignment of conditions to sentences was counter-balanced across participants with a Latin-square design. The experiment always started with three standard trials to establish the sinewave tone as the standard sound. The remaining trials were then presented in a different pseudo-random order for each participant.

**Apparatus**

Participants' eye-movements were recorded with an Eyelink 1000 eye-tracker at 1000 Hz. Viewing was binocular, but only the right eye was recorded. Participants' head was stabilized with a chin-and-forehead rest to reduce head-movements. The stimuli were displayed on a Cambridge Research Systems LCD++ monitor (resolution: 1920 x 1080 pixels; refresh rate: 120 Hz). The sound stimuli were played on a Creative Labs Sound Blaster X-Fi SB0770 sound card and were presented binaurally through Bose QuietComfort 25 noise-cancelling headphones at a 65 dB(A) SPL.



The experiment was programmed in Matlab R2014a (MathWorks, 2014) using the Psychophysics Toolbox v.3.0.11 (Brainard, 1997; Pelli, 1997) and Eyelink libraries (Cornelissen, Peters, & Palmer, 2002). The sound stimuli were played using the low-latency mode of the Psychophysics Toolbox (output latency was 14 ms). The sentences were formatted in a monospaced Courier New 18pt. font and appeared on a single line in the middle of the screen. The sentences were presented with a 50-pixel offset from the left side of the screen and appeared as black text over white background. The letter width was 14 pixels and the eye-to-screen distance was 80 cm. At this distance, each letter subtended ~0.34º of visual angle. The experiment was run on a PC in a Windows 7 64-bit environment.

**Procedure**

Participants were tested individually in a session that lasted about 40 minutes. Before the start of the experiment, a 3-point horizontal calibration was performed. Calibration accuracy was then monitored with a drift check presented before each trial and participants were recalibrated whenever necessary (the error was kept at < 0.3º throughout the experiment). All beeps during calibration and drift check were turned off. Participants were instructed to ignore any sounds they may hear and to read the sentences for comprehension. The experiment started with six practice items presented in silence. Each trial began with a black gaze box that was centred at the first letter in the sentence. Once a stable fixation inside the gaze box was detected, the box disappeared and the sentence was presented on the screen.

The gaze-contingent sound presentation was implemented by placing in invisible boundary (Rayner, 1975) before each of the five target words (for more details on this technique, see Eiter & Inhoff, 2010; Inhoff, Connine, Eiter, Radach, & Heller, 2004; Inhoff et al., 2002). Once the gaze position of the eye crossed an invisible boundary, the sound was presented.



Depending on the sound onset condition, the sound was played either immediately after crossing the boundary (0 ms delay) or 120 ms after crossing the boundary (120 ms delay). This typically happened as participants made a forward saccade towards the target word. A third of the sentences were followed by a Yes/No comprehension question (see Figure 1f). Participants pressed the left button of the mouse to terminate the trial and to answer the comprehension questions. The questions were used to ensure that participants were reading for comprehension and were not a variable of main interest in the analyses.

**Data Analysis**

The analysis focused on the immediate effect of the sound, which was measured on the first fixation and first saccade after it was played. We did not restrict the analysis only to cases where the Target Word was fixated during first-pass reading, but also included cases where it was skipped and another word was fixated. This was because our hypothesis predicted a general inhibition of saccade planning that should occur regardless of which word is fixated. However, the results did not change when restricting the analysis only to cases where the target word was fixated (see the Supplemental Materials).

The fixation and saccade data were analysed only for sounds played at Target Word locations 2-5 since no novel sounds occurred on Target Word 1. Because the standard sound was presented much more frequently than the novel sounds, only one standard sound was sampled per trial. This sound was picked using the design matrix that was used to counter-balance the presentation of novel sounds. This ensured that a balanced dataset was formed, which had an equal number of standard and novel sounds that were presented equally often on Target Words 2-5. In the fixation data, we analysed only the first fixation duration during which the sound was played. This was because Vasilev et al. (2019) established that the first fixation was the source of



the sound deviance effect. To measure the effect of novel sounds on the execution of the next saccade after the playing the sound, a few measures were used: saccade duration, saccade amplitude, peak and average saccade velocity.

The data were analysed with (Generalised) Linear Mixed Models (GLMM(s)) using the lme4 package v.1.1-21 (Bates, Machler, Bolker, & Walker, 2014) in the R software v.3.6.2 (R Core Team, 2019). The fixed factors in the model were sound type (standard vs novel), sound onset delay (0 vs 120 ms delay), and their interaction. Random intercepts were added for both participants and items (Baayen, Davidson, & Bates, 2008). Additionally, we tried adding random slopes for sound type and sound onset delay (Barr, Levy, Scheepers, & Tily, 2013). However, the models converged only with a random slope for sound onset delay (subjects) for the fixation duration and saccade velocity analyses; and sound type (subjects) for the saccade duration and amplitude analyses. Fixation durations, saccade durations, and saccade amplitude were log-transformed in the models as this improved the distribution of residuals. However, the results did not change when using the untransformed values. Treatment contrast coding was used for both the sound type (baseline: standard) and sound onset delay (baseline: 0 ms delay) conditions. The results were considered statistically significant if the $|t|$- and $|z|$-values were $\geq 1.96$. Additionally, Bayesian LMMs were used to calculate Bayes Factors ($BF_{10}$) to quantify the evidence in support of the null and the alternative hypothesis (Dienes, 2014). Full details are provided in Appendix A. While $BF_{10} > 1$ indicate support for the alternative hypothesis, $BF_{10} < 1$ indicate evidence for the null hypothesis.

Additionally, a survival analysis was done to determine the time course of novelty distraction by estimating the earliest point in time when the effect could be reliably detected. This was done using the confidence interval divergence point analysis (CI-DPA; Reingold & Sheridan,



2014, 2018). This is a survival analysis technique that helps determine the earliest point in time where the distributions of the two sound conditions begin to significantly diverge from one another. The analysis was done with 10 000 bootstrap iterations following the method described in Reingold and Sheridan (2018).

## Results

All participants exhibited a comprehension accuracy score greater than 84.6%, thus indicating they had no problems understanding the sentences. Participants' mean comprehension accuracy was 95.7% in the standard ($SD$= 20.3%) and 95% in the novel sound condition ($SD$= 21.7%). There was no significant difference between the sound conditions, $z$= -1.01. When asked after the experiment, only four participants had some awareness that the timing of the sounds depended on their eye-movements[2]. During the data pre-processing, 6.02% of trials were removed due to eye blinks. Additionally, 5.47% of trials were excluded due to boundary "hooks"[3] and a further 12.19% of trials were excluded because the trigger to play the sound was not sent to the soundcard before the onset of the next fixation[4]. Finally, trials with fixation durations shorter than 80 ms or longer than 1000 ms (0.31%), peak saccade velocity greater than 1000 °/s (0.06%), or saccades with amplitude greater than 15° (0.03%) were discarded as outliers. This left 75.91% of the data for analysis (5825 trials). The distribution of trials per condition was: 1474 (standard –

---

[2] Three participants were aware that the sound presentation depended on their reading speed, although they did not know that the sounds were played only on certain words in the sentence. The fourth participant was aware that the sounds are played on specific words, but they were not certain which the exact words in the sentence were.

[3] A "hook" occurs when participant's gaze drifts across the invisible boundary, thus triggering it, but then crosses it back and the next fixation occurs left of the boundary.

[4] This was done to exclude cases where the sound would be played too late relative to fixation onset. Due to a sound output latency of 14 ms in our system, this ensured that the maximum possible onset of the sound would be 14 ms after fixation onset. However, because the trigger to play the sound was usually sent during the preceding saccade, the sound output latency relative to fixation onset was usually much less (M= 4.28 ms in the 0 ms condition; M= 123.6 ms in the 120 ms delay condition [the additional 120 ms are due to the experimental delay manipulation]).



0 ms delay), 1475 (novel – 0 ms delay), 1452 (standard – 120 ms delay), 1424 (novel – 120 ms delay). There was no significant difference in the number of trials per condition, $\chi^2(1)= 0.128$, $p= 0.72$. First-pass skipping of the target occurred on 10.4% of trials, but these trials were retained as mentioned above.

**First Fixation Duration**

Descriptive statistics for the first fixation duration during which the sound was played are presented in Table 1 and are illustrated in Figure 2. The results from the LMM analysis are presented in Table 2. Consistent with **H1**, there was a main effect of Sound, indicating that novel sounds led to longer fixation durations compared to the standard sound ($d= 0.159$). While there was no main effect of Onset Delay, the interaction between Sound and Onset Delay was significant. This interaction supports **H2** in showing that novel sounds yielded more distraction in the 120 ms delay condition ($d= 0.233$) than in the 0 ms delay condition ($d= 0.09$), which can be clearly seen in Figure 2. The $BF_{10}$ results provided converging evidence to the same conclusion. While the data favoured the null hypothesis of no difference for Onset Delay, it favoured the alternative hypothesis of a true difference for both the main effect of Sound and the interaction between Sound and Onset Delay.

A simple-effects analysis (Lenth, Singmann, Love, Buerkner, & Herve, 2019) of the interaction with Tukey $p$-value adjustments confirmed that the novelty distraction effect (novel – standard difference) was significant in both the 0 ms ($b= -0.033$, $SE= 0.011$, $t= -3.107$, $p= 0.0102$) and the 120 ms delay condition ($b= -0.074$, $SE= 0.011$, $t= - 6.992$, $p <.0001$). However, novel sounds were more distracting in the 120 ms than in the 0 ms delay condition ($b= -0.040$, $SE= 0.011$, $t= -3.513$, $p= 0.0031$), whereas the standard sound did not differ between the two



delay conditions ($b$= 0.002, $SE$= 0.011, $t$= 0.145, $p$= 0.9989). In summary, novel sounds were more distracting when played with a 120 ms delay.

Table 1

*Mean Descriptive Statistics for the First Fixation Duration During Which the Sound is Played and the First Saccade Immediately after Playing the Sound (SDs in Parentheses)*

| Sound | Onset delay | First fixation duration (ms) | Next saccade after playing the sound | | | |
| --- | --- | --- | --- | --- | --- | --- |
| | | | Saccade duration (ms) | Saccade amplitude (º) | Peak saccade velocity (º/s) | Average saccade velocity (º/s) |
| Novel | 0 ms | 242 (79) | 21.5 (6.5) | 2.7 (1.29) | 225 (75.7) | 118 (30.2) |
| Standard | 0 ms | 235 (75.6) | 21.3 (6.2) | 2.7 (1.28) | 221 (74.6) | 117 (30.2) |
| Novel | 120 ms | 255 (89.7) | 22 (6.8) | 2.84 (1.36) | 232 (78) | 121 (30.8) |
| Standard | 120 ms | 235 (78.6) | 21.5 (6.2) | 2.73 (1.22) | 223 (72.5) | 119 (29.1) |

*Note*: 1º = 2.9 letters.

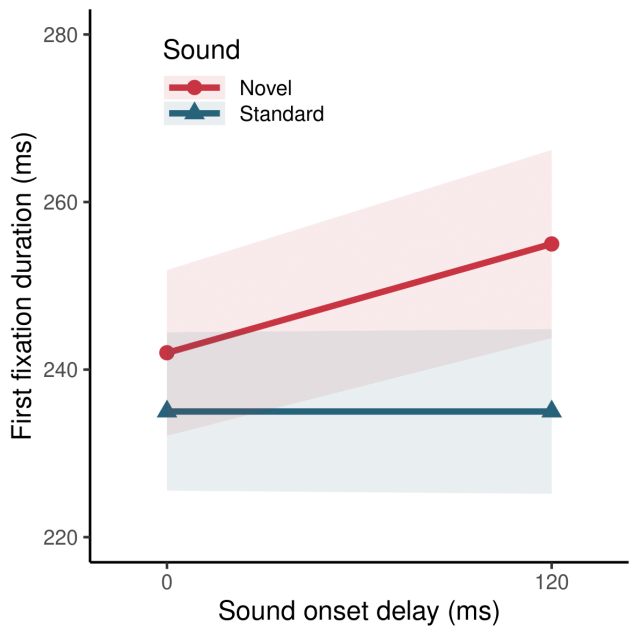

*Figure 2*. Duration of the first fixation during which the sound is played. Shading indicates ± 1 SE.



One possibility is that novel sounds may have affected word processing beyond the first fixation. If participants were more likely to make additional first-pass fixations on the target word, this may indicate lexical processing difficulty. To examine this, we did a post-hoc analysis of first-pass re-fixation probability on the target word. The results (presented in the Appendix B) indicated no significant differences in this measure. This suggests that the effect of novel sounds was largely constrained to the first fixation during which they are played.

Table 2

*LMM Results of the First Fixation Duration during which the Sound is Played*

| Fixed effects | Estimate | Std. Error | t value | $BF_{10}$ |
|---|---|---|---|---|
| Intercept | 5.413 | 0.014 | **373.86** | |
| Sound (Novel vs Standard) | 0.033 | 0.011 | **3.108** | **24.07** |
| Delay (120 vs 0ms) | -0.002 | 0.011 | -0.145 | **0.137** |
| Sound x Delay | 0.042 | 0.015 | **2.792** | **7.88** |
| Random effects | Variance | SD | Corr. | |
| Intercept (item) | 0.00113 | 0.0336 | | |
| Intercept (subjects) | 0.00927 | 0.0963 | | |
| Delay (subjects) | 0.00113 | 0.0336 | -0.09 | |
| Residual | 0.08107 | 0.2847 | | |

*Note*: $BF_{10}$: Bayes factor (values >1 indicate evidence for the alternative hypothesis; values <1 indicate evidence for the null hypothesis). Statistically significant *t*-values (≥1.96) and $BF_{10}$s meeting the 3/ 0.33 threshold (approx. corresponding to the 0.05 alpha level) are formatted in bold.



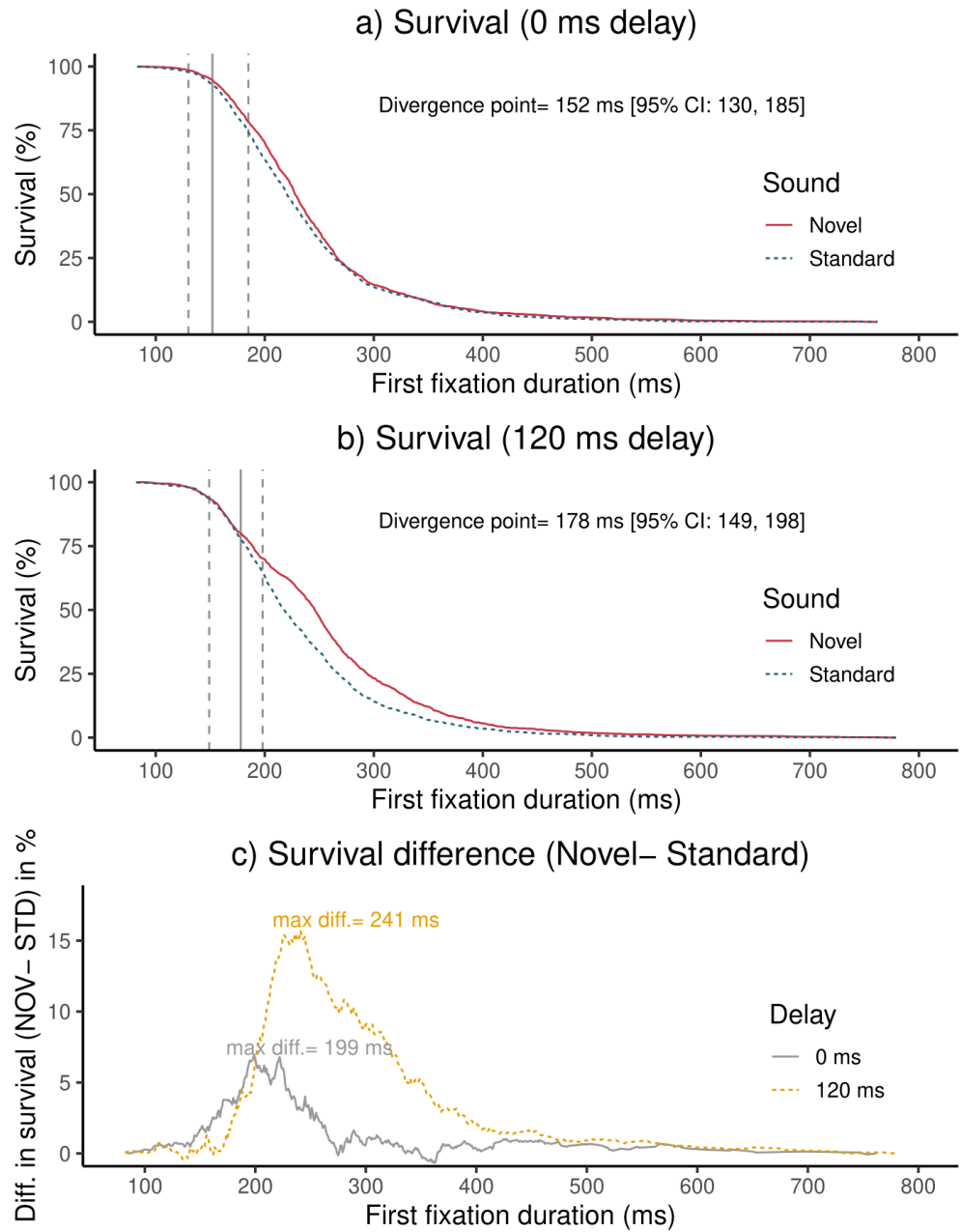

*Figure 3.* Results from the divergence point analysis of the first fixation duration during which the sound is played (**a-b**) and difference in survival between the novel and standard conditions (**c**). In panels (**a-b**), the survival curves are plotted for the two sound conditions, denoting the percentage of "surviving" (i.e., remaining) fixations at each duration bin. The divergence point is illustrated by a vertical solid line and the 95% CI is illustrated by vertical dotted lines. Panel (**c**) shows the difference in "survival" curves between the standard and novel conditions, along with the point of maximum difference (i.e., when the effect was strongest in the distribution of fixations).



To determine the earliest point in time when the novel sound started to significantly influence fixation durations, the CI-DPA analysis (Reingold & Sheridan, 2018) was used with the first fixation data. The results are illustrated in Figure 3. In the 0 ms delay condition, the distributions of the two sound conditions began to significantly diverge at 152 ms [95% CI: 130, 185]. In contrast, the two distributions in the 120 ms onset delay condition started to diverge at 178 ms [95% CI: 149, 198] (58 ms post the sound onset). Therefore, while the divergence point for the 120 ms delay condition occurred some 26 ms later, the novel sounds in both conditions had a relatively quick influence on fixation durations. Nevertheless, as Figure 3c shows, the strongest reduction in terminated fixations between the novel and the standard sound did not occurr until 199 ms in 0 ms delay condition and 241 ms in the 120 ms delay condition (199 and 121 ms post the sound onset, respectively). Therefore, while the difference was detectible statitically early on, it did not reach its greatest magnitude immediately in the distribution of fixation durations.

**Saccade Execution**

The descriptive statistics for saccade measures are shown in Table 1 and the LMM results are presented in Tables 3 and 4. Contrary to **H3.1** and **H3.2**, there were no significant differences between the two sound conditions and no interactions with sound Onset Delay. As Figure 4 shows, novel sounds had no influence on the main sequence of the next saccade. The $BF_{10}$ results showed substantial evidence in favour of the null hypothesis in all but two cases: the main effect of Sound in peak saccade velocity and the Sound x Onset Delay interaction in saccade amplitude. However, even in those two cases, the data still supported the null over the alternative hypothesis. It should be noted that the frequentist model showed a hint of a main effect of Sound in peak saccade velocity (t= 1.89), which was *opposite* to predictions (i.e., peak velocity *increasing* in the



novel compared to the standard sound). However, the Bayesian model still weakly favoured the null hypothesis of no difference (BF$_{10}$ = 0.462). Therefore, while the support for the null was not strong, there was no evidence to support the alternative either. Finally, a post-hoc analysis indicated that the results were generally not modulated by whether participants made an intra-word or an inter-word saccade (see the Supplementary files). In summary, the results suggest that, once triggered, the execution of a saccade was generally not perturbed by the recent presentation of a novel sound.

Table 3

*LMM Results for the Duration and Amplitude of the Next Saccade After Playing the Sound*

| Fixed effects | Saccade duration | | | | Saccade amplitude | | | |
|---|---|---|---|---|---|---|---|---|
| | Estimate | Std. Error | t value | BF$_{10}$ | Estimate | Std. Error | t value | BF$_{10}$ |
| Intercept | 3.013 | 0.016 | **189.77** | | 0.865 | 0.025 | **34.001** | |
| Sound (Novel vs Standard) | 0.005 | 0.011 | 0.472 | **0.062** | -0.006 | 0.02 | -0.279 | **0.140** |
| Delay (120 vs 0ms) | 0.012 | 0.011 | 1.072 | **0.096** | 0.021 | 0.018 | 1.152 | **0.252** |
| Sound x Delay | 0.016 | 0.015 | 1.059 | **0.129** | 0.035 | 0.026 | 1.359 | 0.433 |
| Random effects | Variance | SD | Corr. | | Variance | SD | Corr. | |
| Intercept (item) | 0.00063 | 0.02520 | | | 0.00244 | 0.04937 | | |
| Intercept (subjects) | 0.01209 | 0.10997 | | | 0.02935 | 0.17131 | | |
| Sound (subjects) | 0.00053 | 0.02308 | 0.62 | | 0.00528 | 0.07267 | 0.15 | |
| Residual | 0.08451 | 0.29071 | | | 0.24474 | 0.49471 | | |

*Note*: BF$_{10}$: Bayes factor (values >1 indicate evidence for the alternative hypothesis; values <1 indicate evidence for the null hypothesis). Statistically significant *t*-values (≥1.96) and BF$_{10}$s meeting the 3/ 0.33 threshold (approx. corresponding to the 0.05 alpha level) are formatted in bold.



Table 4

*LMM Results for the Peak and Average Saccade Velocity of the Next Saccade After Playing the Sound*

| Fixed effects | Peak saccade velocity | | | | Average saccade velocity | | | |
|---|---|---|---|---|---|---|---|---|
| | Estimate | Std. Error | t value | BF$_{10}$ | Estimate | Std. Error | t value | BF$_{10}$ |
| Intercept | 219.92 | 5.495 | **40.02** | | 117.103 | 1.785 | **65.604** | |
| Sound (Novel vs Standard) | 4.382 | 2.319 | 1.89 | 0.462 | 0.646 | 1.005 | 0.643 | **0.041** |
| Delay (120 vs 0ms) | 3.196 | 2.59 | 1.234 | **0.190** | 1.326 | 1.049 | 1.264 | **0.076** |
| Sound x Delay | 4.488 | 3.301 | 1.36 | **0.272** | 1.869 | 1.43 | 1.307 | **0.114** |
| Random effects | Variance | SD | Corr. | | Variance | SD | Corr. | |
| Intercept (item) | 49.47 | 7.033 | | | 10.364 | 3.219 | | |
| Intercept (subjects) | 1733.38 | 41.634 | | | 165.983 | 12.883 | | |
| Delay (subjects) | 81.09 | 9.005 | -0.02 | | 5.177 | 2.275 | -0.31 | |
| Residual | 3932.9 | 62.713 | | | 740.827 | 27.218 | | |

*Note*: BF$_{10}$: Bayes factor (values >1 indicate evidence for the alternative hypothesis; values <1 indicate evidence for the null hypothesis). Statistically significant *t*-values (≥1.96) and BF$_{10}$s meeting the 3/ 0.33 threshold (approx. corresponding to the 0.05 alpha level) are formatted in bold.

## Discussion

The present study tested whether the increase in fixation durations during reading in response to unexpected sounds reflects a perturbation of the planning of the next saccade. In addition, it examined whether novels sounds affect not only the planning, but also the execution of the next saccade. Consistent with previous results (Vasilev, Parmentier, et al., 2019), novel sounds led to longer first fixation durations compared to standard sounds immediately after their presentation. Although the magnitude of the effect was small, it was comparable to that of other



auditory distraction effects during reading (Vasilev, Kirkby, & Angele, 2018; but see also Hyönä & Ekholm, 2016; Yan, Meng, Liu, He, & Paterson, 2018). Critically, however, novel sounds were more distracting when presented with a 120 ms delay compared to a 0 ms delay. Because the sound in the 120 ms delay condition was played temporarily closer to when readers usually plan their next saccade, the present results support the SIH, which stated that the observed distraction is due to the transient inhibition of saccade planning.

Interestingly, however, saccade execution parameters such as amplitude, duration, and velocity remained unaffected by novel sounds in both delay conditions. The Bayes Factor analyses favoured the null hypothesis of no difference over the alternative hypothesis, and the evidence for the null was substantial (Jeffreys, 1961) in all but two cases. This suggests that novel sounds affect only the planning but not the execution of saccades (at least when the sound is played during a fixation). Therefore, while novelty distraction was rapid enough to delay the onset of the next saccade, its effect was short-lived and did not affect the movement of the eyes during a saccade once this movement was triggered. In this sense, the present results suggest that novel sounds do not lead to a general inhibition of all oculomotor behaviour but, rather, to a transient one that affects only saccade programming (at least when the sounds are presented up to 120 ms into the fixation).



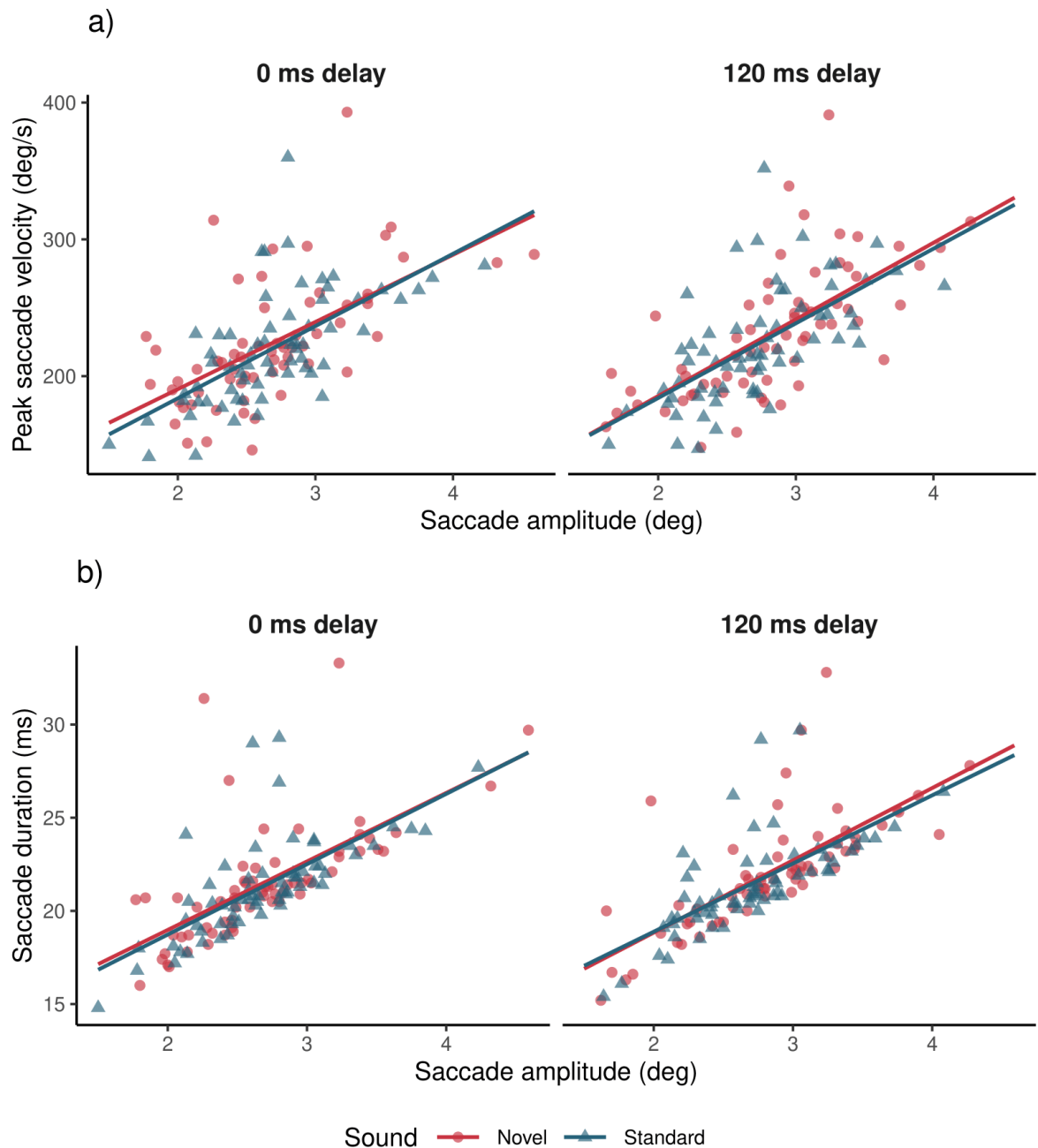

*Figure 4*. A scatterplot of peak saccade velocity and saccade amplitude (**a**) and saccade duration and saccade amplitude (**b**) in the two onset delay conditions ("main sequence"; Bahill et al., 1975). Each data point shows the mean value for novel (circle shape) and standard sounds (triangle shape) for each subject. Lines show the regression slopes.



The greater distraction in first fixations durations in the 120 ms delay condition is consistent with the notion that unexpected sounds lead to a general and temporary inhibition of motor responses (Dutra et al., 2018; Wessel, 2017; Wessel & Aron, 2013). However, such inhibition would be expected to affect all oculomotor behaviour, including the execution of saccades. Because saccadic variables such as peak velocity are assumed to be closely correlated with the firing of brainstem neurons (e.g., Di Stasi, Catena, Cañas, Macknik, & Martinez-Conde, 2013; Fuchs et al., 1985), our results suggest that novel sounds do not affect the neural circuits that control the movement of the eyes during a saccade. While this result may appear at odds with Wessel and Aron's (2013) global motor suppression account, this suppression could be very time-sensitive and disappear by the time the eyes are in motion. In fact, Wessel and Aron (2013) reported that inhibition of motor-evoked potentials was present 150 ms after the sound onset, but had already disappeared 25 ms later. Therefore, the inhibition in the present experiment may be too transient to affect the execution of the next saccade.

A stronger test of whether novel sounds inhibit saccade execution would be to play them during a saccade. However, as most saccades are considerably shorter than the 60 to 150 ms needed for the effect to occur, the influence of novel sounds on saccade execution may be better studied in smooth pursuit tasks where observers have to continuously follow a moving object with their eyes (e.g., Lisberger, Morris, & Tychsen, 1987; Robinson, 1965). Of course, one can also try to adjust the timing of the sounds in the current paradigm to maximise the chance that the



strongest inhibition occurs exactly during the saccade. However, as reading saccades are very short, this critical point can be easily missed. As such, simple saccade tasks where there is strong experimental control over when the next saccade occurs may be preferable.

The time-course of novelty distraction in first fixation durations revealed that novel sounds first began to have an effect some 150 ms after fixation onset in the 0 ms delay condition, and some 180 ms after fixation onset in the 120 ms delay condition. The 150 ms onset in the former condition is in line with Wessel and Aron's (2013) findings. Interestingly, the distraction effect in the 120 ms delay condition first started some 60 ms after the sound's onset (i.e., 180-120= 60 ms). Therefore, novel sounds exerted a faster effect on fixation durations when the sound was played, on average, during the second half of fixations. This rapid effect echoes Graupner et al.'s (2007) results where an auditory deviant caused a "first" saccadic inhibition some 90 ms after sound onset (the sounds in their study were played 100 ms after fixation onset, making it similar to the 120 ms delay condition in our study). Additionally, the early onset in the 120 ms condition is generally consistent with Widmann et al.'s, (2014) finding that microsaccadic inhibition originating some 80-100 ms post the sound onset may indicate the fast categorisation of auditory sounds.

Nevertheless, it is important to emphasize that the divergence point analysis indicates only the *first* point in time where a difference in the fixation duration distributions can be detected. In fact, as Figure 3c shows, the maximum difference between the two sound conditions



did not occur until ~ 200 ms after the sound presentation in the 0 ms delay condition and ~ 120 ms after the sound presentation in the 120 ms delay condition (i.e., 240-120= 120 ms). Therefore, despite the early onset, the effect needed some 50-60 ms to reach its maximum magnitude, and fixations continued to be affected well beyond the initial divergence point.

The earlier onset of the effect in the 120 ms delay condition may occur because novel sounds usually coincided with the critical period during which saccade planning takes place. Because very few fixations terminated within the first 100 ms from the sound's onset in the 0 ms delay condition, such early inhibition may not be seen in that condition[5]. Therefore, the overall pattern of inhibition was somewhat similar across the two delay conditions, but it was much stronger and started earlier in the 120 ms delay condition. Thus, while novelty distraction was present in both delay conditions, the magnitude of the effect was greater in the 120 ms condition because it affected a larger proportion of fixations.

What is the purpose of this temporary inhibition of eye-movement control by novel sounds? It may represent an adaptive response in the face of an unexpected contextual change where readers delay their oculomotor plans to react to something unexpected (i.e., auditory novelty pointing towards potential danger). Unexpected sounds are arguably surprising events (e.g., Parmentier et al., 2019; Wessel et al., 2016). As such, this "freezing" of oculomotor plans prior to goal-directed saccades may allow readers time to process this contextual change and select an appropriate response (see also Wessel, 2017, 2018). Generally speaking, sounds violating sensory prediction can be regarded as one instance of an unexpected change in our immediate environment, which brings some uncertainty and calls for a reappraisal of current actions plans. In line with this contention, novel sounds have been shown to render the repetition of one's behaviour more difficult to facilitate the adoption of a different action plan (Bendixen & Schröger, 2008; Parmentier, 2016; Roeber, Berti, Müller, Widmann, & Schröger, 2009; Roeber, Berti, Widmann, & Schröger, 2005). Interestingly, there is some evidence that saccadic inhibition can also occur in response to visual deviants (e.g., Godijn & Kramer, 2008; Graupner et al., 2007). Therefore, one interesting question for future research would be to explore whether this inhibitory response is specific to auditory stimuli, or if it reflects a more general response to unexpected contextual changes, which also extends to the visual domain.

---

[5] Following the typical convention in reading research, fixations shorter than 80 ms were discarded before analysis. However, the results from the survival analysis did not change when these fixations were retained in the data (the divergence point in the 0 ms delay condition was then 150 ms [95% CI: 122, 184].

Currently, it is not clear whether the increase in fixation durations simply reflects a neural inhibition that slows down the neural firing leading up to a saccade or if the current saccade plan is cancelled in order to plan a different one. Computational models of reading (e.g., Engbert et al., 2005; Reichle et al., 1998) assume that saccade planning occurs in two stages: one cancellable and another non-cancellable. This proposition is based on Becker and Jürgens' (1979) findings that saccades can be programmed in parallel and modified if a new target is presented sufficiently early in the planning stages. In the context of our study, the next saccade plan may be reset or



cancelled due to the detection of auditory novelty and, perhaps, the following reappraisal of action plans, before the saccade is re-programmed. Alternatively, saccade plans may stay the same, but simply take longer to program because they are interrupted by a transient global neuro-motor inhibition. Future research might possibly adjudicate between these two explanations by using the double-step paradigm (e.g., Becker, 1972), where a target jumps to a new location prior to the execution of a saccade towards it. If novel sounds affect saccadic plans, then they should also influence the parallel planning of saccades in response to double-step stimuli.

Given that the maximum inhibition in the 120 ms delay condition occurred close to when the "average" fixation duration in the experiment was ending and the next saccade was starting (~ 240 ms; see Figure 3c), it is perhaps surprising that novel sounds did not affect saccade execution. In the E-Z Reader framework, the observed inhibition was well within the last 25-50 ms of the fixation, corresponding to the non-labile stage where the direction and distance of the next saccade is communicated to the motor system (Reichle et al., 2003, 2009). However, the subsequent movement of the eyes during the saccade was not affected. We speculate that this may occur because the inhibitory effect is very time-sensitive and may affect behaviour only within a very narrow window of time when the novel sound is first detected (~ 25 ms in Wessel & Aron, 2013). Additionally, saccades are very quick ballistic movements- their duration and velocity are not under voluntary control (Leigh & Zee, 1999) and they are usually too short to be influenced by perceptual feedback (Optican & Pretegiani, 2017). In this sense, there may be little



adaptive value in inhibiting a saccadic movement once it has already started because the

individual will have a limited ability to act upon the unexpected stimulus during the saccade. As

this is the first study to look at this topic, more evidence is required to reach a firmer conclusion.

Finally, while the present data is consistent with the SIH, it does not prove that novelty

distraction is completely independent of language processing. For instance, the experimental

manipulation is not sensitive to higher-level syntactic or integration processes that may occur

after readers have left the fixated word. While the available data suggests that distraction is

mostly constrained to the current fixation (Vasilev, Parmentier, et al., 2019), further research is

needed to better understand if unexpected sounds can have any effect at all on language

processing. Nevertheless, we anticipate that the present results will be useful in stimulating

further research on how unexpected sounds may affect complex everyday tasks such as reading.

In summary, the present study showed that novel, task-irrelevant sounds exert greater

distraction in a reading task when they occur temporarily close to the end of fixations, thus

suggesting that they disrupt the planning of the next saccade. At the same time, novel sounds do

not appear to influence the execution of the next saccade. This suggests that their inhibition is

short-lived, and the distractive impact of the novel sounds generally does not extend to

subsequent oculomotor behaviour. These results provide further evidence that unexpected sounds

can lead to a rapid inhibition of oculomotor planning in everyday tasks such as reading.



## Acknowledgments

These data were presented at the 2020 EPS meeting (8-10 January) in London, UK. We thank Rachel Sauvarin for her help with some of the data collection. We are also grateful to 3 anonymous reviewers for their valuable comments on a previous version of this manuscript. Data files, analysis scripts and materials are available at: https://doi.org/10.17605/OSF.IO/Q8HJ9

## Declaration of Conflicting Interests

The Authors declare that there is no conflict of interest.

## Funding

Martin Vasilev was supported by a doctoral and a post-doctoral fellowship from Bournemouth University. Fabrice Parmentier was supported by a research grant from the Spanish Ministry of Science, Innovation and Universities (PSI2014-54261-P), the Spanish State Agency for Research (AEI), and the European Regional Development Fund (FEDER).



Appendix A

**Bayesian Linear Mixed Models**

To calculate the evidence for or against the null hypothesis, supplementary Bayesian LMMs were fitted. For consistency, they had the same model structure as the frequentist LMMs reported in the main text. The Bayesian LMMs were fitted with the *brms* R package v.2.13.5 (Bürkner, 2017, 2018) using the Stan software (Carpenter et al., 2017).

Weakly informative Normal priors were specified based on previous results and general constraints of the oculomotor system (see Table A1 for summary). In the first fixation duration, the expected slope effects were between ~0-20 ms (0-0.08 in log units) based on previous findings (Vasilev, Parmentier, et al., 2019). Therefore, a ~N(0, 0.08) prior was used where 95% of the probability density falls between -0.16 and 0.16 (-40, 40 ms). Saccades are much shorter than fixations, so slope effects of about 5 ms were expected (0.2 in log units); therefore, a ~N(0, 0.2) prior was used where 95% of the probability distribution falls between -0.4 and 0.4 (-10, 10 ms). Auditory distraction effects in saccade amplitude are relatively small (Vasilev, Liversedge, Rowan, Kirkby, & Angele, 2019), so we expected differences of up to a letter (0.15 in log units). Therefore, using a ~N(0, 0.15) prior, 95% of the prior density falls between -0.3 and 0.3 (-0.6, 0.6º, or approx. ± 2 letters). Human saccades of the amplitude typical for reading usually have velocities of 100-300 º/s (Boghen, Troost, Daroff, Dell'Osso, & Birkett, 1974). We expected slope effects of up to 30 º/s; therefore, using a ~ N(0, 30) prior, 95% of the probability density falls between -60 and 60 º/s. For target re-fixation probability, we expected differences of up to 10% (0.75 in logit units); therefore, using a ~ N(0, 0.75) prior, 95% of the probability density falls between -1.5 and 1.5 (-20, 20%). For the model intercepts, reasonable values were used based on the typical averages for each measure (see Table A1).



Table A1

*Summary of Priors Used in the Bayesian LMM Models*

| Measure | Priors | | 95% prior probability density of the expected effect |
|---|---|---|---|
| | Intercept | Slopes | |
| First fixation duration | ~N(0, 5) | ~N(0, 0.08) | [-0.16, 0.16 in log units]; ~ [-40, 40 in ms] |
| Saccade duration | ~N(0, 5) | ~N(0, 0.2) | [-0.4, 0.4 in log units]; ~ [-10, 10 in ms] |
| Saccade amplitude | ~ N(0, 5) | ~N(0, 0.15) | [-0.3, 0.3 in log units]; ~ [-0.6, 0.6 in deg] |
| Peak saccade velocity | ~ N(0, 125) | ~ N(0, 30) | [-60, 60 in deg/s] |
| Average saccade velocity | ~ N(0, 125) | ~ N(0, 30) | [-60, 60 in deg/s] |
| Target re-fixation prob. | ~ N(0, 3) | ~ N(0, 0.75) | [-1.5, 1.5 in logit unit]; ~ [-0.20, 0.20 in prob.] |

*Note*: differences in log units between the measures reflect differences in the intercept and the non-linear nature of the log function.

Sampling from the posterior distribution was done with 10 Markov chains with 6000 iterations each. The first 1000 iterations were discarded as burn-in. Therefore, the summary of the posterior was based on 50 000 samples. Bayes factors ($BF_{10}$) were calculated with the *brms* package using the Savage-Dickey density ratio method (Dickey & Lientz, 1970; Morey, Rouder, Pratte, & Speckman, 2011). While $BF_{10}$ >1 indicates evidence in support of the alternative hypothesis, $BF_{10}$ <1 indicate evidence in support of the null hypothesis. $BF_{10}$ of 1 indicates that the model cannot distinguish between the two hypotheses. According to Jeffreys (1961), $BF_{10}$ of 3 or 1/3 indicates substantial evidence in support of the alternative or null hypothesis, respectively. These cut-offs were originally designed to roughly correspond to Fisher's (1937) 0.05 significance threshold. However, Bayesian evidence is continuous in nature and can be



interpreted without any cut-offs. The results from the models are reported in Tables A2-A4. The

BF$_{10}$ results are reported in the main text along with the frequentist results for easier comparison.

Table A2

*Bayesian LMM Results for the First Fixation Duration during which the Sound is Played and First-pass Re-fixation Probability of the Target Word*

| Fixed effects | Fixation duration | | | First-pass refixation probability | | |
|---|---|---|---|---|---|---|
| | Estimate | Std. Error | 95% CrI | Estimate | Std. Error | 95% CrI |
| Intercept | 5.412 | 0.015 | **[5.383, 5.442]** | -1.824 | 0.114 | **[-2.053, -1.604]** |
| Sound (Novel vs Standard) | 0.033 | 0.010 | **[0.013, 0.053]** | 0.060 | 0.109 | [-0.155, 0.275] |
| Delay (120 vs 0ms) | -0.001 | 0.011 | [-0.023, 0.020] | -0.116 | 0.117 | [-0.346, 0.112] |
| Sound x Delay | 0.041 | 0.015 | **[0.012, 0.069]** | -0.265 | 0.161 | [-0.581, 0.048] |
| Random effects | Variance | SD | Corr. | Variance | SD | Corr. |
| Intercept (items) | 0.00112 | 0.03349 | | 0.05745 | 0.23969 | |
| Intercept (subjects) | 0.00970 | 0.09851 | | 0.35328 | 0.59437 | |
| Delay (subjects) | 0.00066 | 0.02571 | -0.012 | | | |
| Delay (items) | | | | 0.05457 | 0.23361 | -0.319 |
| Sound (subjects) | | | | | | |
| Residual | 0.08124 | 0.28503 | | | | |

*Note*: CrI: Credible intervals. Effects where the 95% Credible Interval excludes 0 are formatted in bold.



Table A3

*Bayesian LMM Results for the Peak and Average Saccade Velocity of the Next Saccade After Playing the Sound*

| Fixed effects | Peak saccade velocity | | | Average saccade velocity | | |
|---|---|---|---|---|---|---|
| | Estimate | Std. Error | 95% CrI | Estimate | Std. Error | 95% CrI |
| Intercept | 219.5 | 5.637 | **[208.5, 230.7]** | 117.12 | 1.821 | **[113.5, 120.7]** |
| Sound (Novel vs Standard) | 4.380 | 2.303 | [-0.135, 8.905] | 0.634 | 0.996 | [-1.308, 2.580] |
| Delay (120 vs 0ms) | 3.191 | 2.551 | [-1.838, 8.193] | 1.316 | 1.037 | [-0.717, 3.346] |
| Sound x Delay | 4.467 | 3.268 | [-1.929, 10.87] | 1.881 | 1.419 | [-0.910, 4.677] |
| Random effects | Variance | SD | Corr. | Variance | SD | Corr. |
| Intercept (items) | 49.0759 | 7.00542 | | 10.3540 | 3.21791 | |
| Intercept (subjects) | 1807.83 | 42.5185 | | 170.502 | 13.0576 | |
| Delay (subjects) | 59.9086 | 7.74007 | 0.016 | 3.0212 | 1.73816 | -0.217 |
| Residual | 3940.57 | 62.7739 | | 741.799 | 27.2359 | |

*Note*: CrI: Credible intervals. Effects where the 95% Credible Interval excludes 0 are formatted in bold.



Table A4

*Bayesian LMM Results for the Duration and Amplitude of the Next Saccade After Playing the Sound*

| Fixed effects | Saccade duration | | | Saccade amplitude | | |
|---|---|---|---|---|---|---|
| | Estimate | Std. Error | 95% CrI | Estimate | Std. Error | 95% CrI |
| Intercept | 3.013 | 0.016 | **[2.981, 3.046]** | 0.865 | 0.026 | **[0.814, 0.916]** |
| Sound (Novel vs Standard) | 0.005 | 0.011 | [-0.017, 0.027] | -0.005 | 0.020 | [-0.045, 0.034] |
| Delay (120 vs 0ms) | 0.012 | 0.011 | [-0.009, 0.033] | 0.021 | 0.018 | [-0.014, 0.057] |
| Sound x Delay | 0.016 | 0.015 | [-0.014, 0.046] | 0.035 | 0.025 | [-0.015, 0.084] |
| Random effects | Variance | SD | Corr. | Variance | SD | Corr. |
| Intercept (items) | 0.00058 | 0.02401 | | 0.00238 | 0.04874 | |
| Intercept (subjects) | 0.01311 | 0.11451 | | 0.03116 | 0.17651 | |
| Sound (subjects) | 0.00056 | 0.02377 | 0.435 | 0.00451 | 0.06712 | 0.204 |
| Residual | 0.08459 | 0.29085 | | 0.24517 | 0.49515 | |

*Note*: Effects where the 95% Credible Interval excludes 0 are formatted in bold.



Appendix B

**First-pass Re-fixation Probability of the Target Word (Post-Hoc Analysis)**

First-pass re-fixation probability was analysed to determine if the processing of the target word was disrupted beyond the first fixation. Descriptive statistics are shown in Table B1 and the GLMM results are shown in Table B2. There were no significant differences across the conditions, which suggests that novel sounds did not disrupt reading beyond the initial fixation during which the sound was played. Therefore, there was no evidence to suggest that target word processing was affected and that readers needed to make additional fixations to process the target word. The $BF_{10}$ analysis was consistent with the lack of difference, except for the Sound x Delay interaction where the $BF_{10}$ was close to 1 and generally could not distinguish between the null and alternative hypothesis. Examining the means indicated that there was a 0.6% increase in refixation probability with novel sounds in the 0 ms delay condition, but that there was a 2.4 % *decrease* in re-fixation probability with novel sounds in the 120 ms delay condition. Therefore, this trend is opposite to predictions, as novel sounds would be expected to increase re-fixation probability, not decrease it, in the 120 ms condition. Clearly, the re-fixation probability data cannot explain the pattern of results in first fixation durations. If anything, re-fixation probability appeared to decrease slightly in the 120 ms condition.



Table B1

*Mean First-Pass Re-Fixation Probability on the Target Word (SD in Parenthesis)*

| Sound | Onset delay | First-pass re-fixation probability |
|-------|-------------|-----------------------------------|
| Novel | 0 ms | 0.162 (0.369) |
| Standard | 0 ms | 0.156 (0.363) |
| Novel | 120 ms | 0.120 (0.325) |
| Standard | 120 ms | 0.144 (0.351) |

Table B2

*GLMM Results for First-Pass Re-Fixation Probability on the Target Word*

| Fixed effects | Estimate | Std. Error | t value | $BF_{10}$ |
|---------------|----------|------------|---------|-----------|
| Intercept | -1.834 | 0.114 | **-16.15** | |
| Sound (Novel vs Standard) | 0.067 | 0.111 | 0.609 | **0.168** |
| Delay (120 vs 0ms) | -0.091 | 0.121 | -0.749 | **0.263** |
| Sound x Delay | -0.277 | 0.164 | -1.686 | 0.823 |
| Random effects | Variance | SD | Corr. | |
| Intercept (items) | 0.09619 | 0.3102 | | |
| Intercept (subjects) | 0.32480 | 0.5699 | | |
| Delay (items) | 0.09299 | 0.3049 | -0.77 | |

*Note*: Statistically significant *t*-values are formatted in bold.



<div align="center">References</div>

## Supplemental files

### Breakdown of the Next Saccade by Saccade Type (Intra-word vs. Inter-word)

Saccade type (intra-word vs inter-word) was added to the saccade execution analyses to examine whether the results were modulated by this variable[6]. *Intra-word* saccade refers to cases where the next saccade lands on the same word as the previous fixation (i.e., the word is re-fixated). Inter-word saccade refers to cases where the next saccade lands on a different word (i.e., another word is fixated). Inter-word saccades tend to be longer than intra-word saccades as the eyes need to travel a greater distance. Additionally, because saccade velocity increases with saccade amplitude (Bahill, Clark, & Stark, 1975), inter-word saccades should also have higher velocities than intra-word saccades.

It could be argued that novel sounds may be less likely to influence intra-word saccades due to their shorter duration, as the effect may have less time to occur. However, on the other hand, shorter saccades will also be completed closer in time to when the sound was played. Because the motor inhibition effect of novel sounds has been found to be transient and time-sensitive (Wessel & Aron, 2013), this may paradoxically make them more susceptible to disruption as a larger proportion of the saccade may be affected by the sound. Therefore, while these two factors may balance each other out, it is still interesting to test if the results are influenced by the type of saccade that participants made.

The descriptive statistics are shown in Table S1 and the LMM results are shown in Tables S2 and S3. Inter-word saccades had longer duration ($d$= 1.304), greater amplitude ($d$= 1.480), higher peak velocity ($d$= 1.179), and higher average velocity ($d$= 1.502) compared to intra-line saccades. However, Sound Type and Delay Onset again did not have any effect on any of the saccade execution variables. Critically, saccade type (intra-word vs inter-word) also did not modulate the results and there were no significant interactions with the experimental conditions.

---

[6] We thank an anonymous Reviewer for this suggestion.

Table S1

*Mean Descriptive Statistics for the First Saccade Immediately after Playing the Sound, Broken Down by Saccade Type (Intra-word vs Inter-word; SDs in Parentheses)*

| Sound | Onset delay | Saccade duration (ms) | Saccade amplitude (°) | Peak saccade velocity (°/s) | Average saccade velocity (°/s) |
|---|---|---|---|---|---|
| **Intra-word next saccade (N= 831 obs.)** | | | | | |
| Novel | 0 ms | 15.2 (5.48) | 1.37 (0.68) | 157 (55.9) | 85.5 (21.9) |
| Standard | 0 ms | 15.3 (5.26) | 1.46 (0.91) | 159 (58.5) | 87.1 (23.8) |
| Novel | 120 ms | 15.0 (5.18) | 1.36 (0.72) | 152 (57.1) | 83.7 (22.5) |
| Standard | 120 ms | 15.5 (4.95) | 1.46 (0.75) | 161 (60.9) | 87.4 (22.6) |
| **Inter-word next saccade (N= 4993 obs.)** | | | | | |
| Novel | 0 ms | 22.7 (5.99) | 2.96 (1.22) | 237 (72.1) | 124 (27.5) |
| Standard | 0 ms | 22.4 (5.77) | 2.93 (1.21) | 232 (71.6) | 123 (27.8) |
| Novel | 120 ms | 23.0 (6.50) | 3.04 (1.31) | 243 (74.1) | 126 (28.2) |
| Standard | 120 ms | 22.5 (5.77) | 2.94 (1.15) | 234 (69.2) | 124 (26.8) |

*Note*: Intra-word saccades refer to cases where the next saccade after playing the sound lands on the same word as the preceding fixation. Inter-word saccades refer to cases where the next saccade after playing the sound lands on a different word to the right.

1° = 2.9 letters.

The Bayes Factor (BF) results largely supported the same conclusion. The majority of the BFs indicated strong support for the alternative hypothesis (i.e., no effect). In the remaining



cases, all but one effect were more consistent with the null than the alternative hypothesis (excluding the main effect of Saccade Type, which was clearly in line with the alternative). However, the BF for the Sound Type x Onset Delay x Saccade Type interaction in peak saccade velocity was close to 1, meaning that it could distinguish between the null and the alternative hypotheses. Therefore, overall, there was little evidence to suggest that the results were influenced by the type of saccade that participants made.

Table S2

*LMM Results for the Duration and Amplitude of the Next Saccade After Playing the Sound, with Saccade Type (Intra-word vs. Inter-word) Added as a Factor in the Analysis*

| Fixed effects | Saccade duration | | | | Saccade amplitude | | | |
|---|---|---|---|---|---|---|---|---|
| | Estimate | Std. Error | t value | $BF_{10}$ | Estimate | Std. Error | t value | $BF_{10}$ |
| Intercept | 2.685 | 0.022 | **123.5** | | 0.27 | 0.034 | **7.99** | |
| Sound (Novel vs Standard) | -0.007 | 0.024 | -0.305 | **0.121** | -0.044 | 0.04 | -1.108 | 0.473 |
| Delay (120 vs 0ms) | 0.007 | 0.025 | 0.279 | **0.125** | -0.0001 | 0.042 | -0.003 | **0.238** |
| Saccade Type | 0.389 | 0.019 | **20.83** | **$4 \times 10^{18}$** | 0.706 | 0.031 | **22.7** | **$6 \times 10^{18}$** |
| Sound x Delay | -0.016 | 0.036 | -0.454 | **0.188** | -0.01 | 0.06 | -0.162 | 0.339 |
| Sound x Saccade type | 0.018 | 0.026 | 0.676 | **0.159** | 0.051 | 0.043 | 1.169 | 0.555 |
| Delay x Saccade type | -0.002 | 0.027 | -0.077 | **0.132** | 0.011 | 0.045 | 0.243 | **0.266** |
| Sound x Delay x Saccade type | 0.022 | 0.039 | 0.574 | **0.217** | 0.024 | 0.065 | 0.366 | 0.401 |
| Random effects | Variance | SD | Corr. | | Variance | SD | Corr. | |
| Intercept (items) | 0.00089 | 0.0299 | | | 0.00289 | 0.05385 | | |
| Delay (items) | | | | | 0.00075 | 0.02741 | 0.01 | |
| Intercept (subjects) | 0.01092 | 0.1045 | | | 0.01937 | 0.13919 | | |
| Sound (subjects) | | | | | 0.00078 | 0.02789 | 0.47 | |
| Residual | 0.06571 | 0.2563 | | | 0.18169 | 0.42626 | | |





Table S3

*LMM Results for the Peak and* Average Velocity *of the Next Saccade After Playing the Sound, with Saccade Type (Intra-word vs. Inter-word) Added as a Factor in the Analysis*

| Fixed effects | Peak saccade velocity | | | | Average saccade velocity | | | |
|---|---|---|---|---|---|---|---|---|
| | Estimate | Std. Error | t value | BF$_{10}$ | Estimate | Std. Error | t value | BF$_{10}$ |
| Intercept | 157.1 | 6.392 | **24.58** | | 87.62 | 2.199 | **39.84** | |
| Sound (Novel vs Standard) | 4.324 | 5.36 | 0.807 | **0.205** | 0.202 | 2.267 | 0.089 | **0.074** |
| Delay (120 vs 0ms) | 3.556 | 5.637 | 0.631 | **0.203** | 0.636 | 2.364 | 0.269 | **0.077** |
| Saccade Type | 74.52 | 4.17 | **17.87** | **4x 10$^{17}$** | 34.95 | 1.771 | **19.73** | **1x 10$^{15}$** |
| Sound x Delay | -10.28 | 8.053 | -1.276 | 0.490 | -3.018 | 3.427 | -0.881 | **0.160** |
| Sound x Saccade type | 0.628 | 5.81 | 0.108 | **0.184** | 0.794 | 2.473 | 0.321 | **0.085** |
| Delay x Saccade type | -1.832 | 6.034 | -0.304 | **0.190** | 0.13 | 2.56 | 0.051 | **0.083** |
| Sound x Delay x Saccade type | 14.11 | 8.695 | 1.622 | 0.877 | 4.271 | 3.698 | 1.155 | **0.229** |
| Random effects | Variance | SD | Corr. | | Variance | SD | Corr. | |
| Intercept (items) | 69.39 | 8.330 | | | 11.6599 | 3.4147 | | |
| Sound (items) | 44.69 | 6.685 | -0.53 | | | | | |
| Intercept (subjects) | 1642.80 | 40.532 | | | 134.484 | 11.5967 | | |
| Delay (subjects) | 50.83 | 7.130 | 0.08 | | 0.05183 | 0.2277 | -1.00 | |
| Residual | 3233.71 | 56.866 | | | 588.550 | 24.260 | | |





**Analysis after Excluding Target Skips**

Table S4

*Mean Descriptive Statistics for the First Fixation Duration During Which the Sound is Played and the First Saccade Immediately after Playing the Sound (Excluding Target Word Skips; SDs in Parentheses)*

| Sound | Onset delay | First fixation duration (ms) | Next saccade after playing the sound | | | |
|---|---|---|---|---|---|---|
| | | | Saccade duration (ms) | Saccade amplitude (º) | Peak saccade velocity (º/s) | Average saccade velocity (º/s) |
| Novel | 0 ms | 242 (80.1) | 21.5 (6.46) | 2.70 (1.26) | 225 (74.7) | 118 (29.8) |
| Standard | 0 ms | 235 (76.4) | 21.3 (6.18) | 2.69 (1.25) | 222 (74.4) | 118 (29.9) |
| Novel | 120 ms | 255 (91.1) | 22.0 (6.74) | 2.85 (1.29) | 234 (76.9) | 122 (30.2) |
| Standard | 120 ms | 235 (79.4) | 21.5 (6.17) | 2.74 (1.20) | 224 (72.6) | 119 (28.9) |

*Note*: 1º = 2.9 letters.



Table S5

*LMM Results of the First Fixation Duration during which the Sound is Played (with Target Word Skips Excluded from Analysis)*

| Fixed effects | Estimate | Std. Error | t value | $BF_{10}$ |
|---|---|---|---|---|
| Intercept | 5.41 | 0.015 | **368.5** | |
| Sound (Novel vs Standard) | 0.034 | 0.011 | **3.043** | **18.73** |
| Delay (120 vs 0ms) | -0.001 | 0.012 | -0.069 | **0.150** |
| Sound x Delay | 0.039 | 0.016 | **2.406** | **3.293** |
| Random effects | Variance | SD | Corr. | |
| Intercept (items) | 0.00142 | 0.03773 | | |
| Intercept (subjects) | 0.00890 | 0.09434 | | |
| Delay (subjects) | 0.00154 | 0.03921 | -0.03 | |
| Residual | 0.08144 | 0.28538 | | |

*Note*: $BF_{10}$: Bayes factor (values >1 indicate evidence for the alternative hypothesis; values <1 indicate evidence for the null hypothesis). Statistically significant *t*-values (≥1.96) and $BF_{10}$s meeting the 3/ 0.33 threshold (approx. corresponding to the 0.05 alpha level) are formatted in bold.



Table S6

*LMM Results for the Duration and Amplitude of the Next Saccade After Playing the Sound (with Target Word Skips Excluded from Analysis)*

| Fixed effects | Saccade duration | | | | Saccade amplitude | | | |
|---|---|---|---|---|---|---|---|---|
| | Estimate | Std. Error | t value | $BF_{10}$ | Estimate | Std. Error | t value | $BF_{10}$ |
| Intercept | 3.017 | 0.016 | **189.4** | | 0.87 | 0.026 | **33.82** | |
| Sound (Novel vs Standard) | 0.005 | 0.012 | 0.415 | **0.068** | -0.007 | 0.023 | -0.306 | **0.154** |
| Delay (120 vs 0ms) | 0.014 | 0.012 | 1.249 | **0.122** | 0.026 | 0.02 | 1.344 | 0.344 |
| Sound x Delay | 0.014 | 0.016 | 0.879 | **0.120** | 0.034 | 0.028 | 1.234 | 0.383 |
| Random effects | Variance | SD | Corr. | | Variance | SD | Corr. | |
| Intercept (items) | 0.00065 | 0.02544 | | | 0.00271 | 0.05204 | | |
| Intercept (subjects) | 0.01165 | 0.10793 | | | 0.02866 | 0.16930 | | |
| Sound (subjects) | 0.00141 | 0.03754 | 0.31 | | 0.00854 | 0.09243 | 0.05 | |
| Residual | 0.08344 | 0.28886 | | | 0.24048 | 0.49039 | | |

*Note*: $BF_{10}$: Bayes factor (values >1 indicate evidence for the alternative hypothesis; values <1 indicate evidence for the null hypothesis). Statistically significant *t*-values (≥1.96) and $BF_{10}$s meeting the 3/ 0.33 threshold (approx. corresponding to the 0.05 alpha level) are formatted in bold.



Table S7

*LMM Results for the Peak and Average Saccade Velocity of the Next Saccade After Playing the Sound (with Target Word Skips Excluded from Analysis)*

| Fixed effects | Peak saccade velocity | | | | Average saccade velocity | | | |
|---|---|---|---|---|---|---|---|---|
| | Estimate | Std. Error | t value | $BF_{10}$ | Estimate | Std. Error | t value | $BF_{10}$ |
| Intercept | 220.8 | 5.271 | **41.88** | | 117.363 | 1.764 | **66.531** | |
| Sound (Novel vs Standard) | 4.075 | 2.718 | 1.499 | **0.316** | 0.704 | 1.066 | 0.66 | **0.044** |
| Delay (120 vs 0ms) | 3.689 | 2.503 | 1.474 | **0.232** | 1.602 | 1.115 | 1.437 | **0.103** |
| Sound x Delay | 4.769 | 3.552 | 1.343 | **0.273** | 1.702 | 1.523 | 1.117 | **0.093** |
| Random effects | Variance | SD | Corr. | | Variance | SD | Corr. | |
| Intercept (items) | 64.04 | 8.002 | | | 11.825 | 3.439 | | |
| Intercept (subjects) | 1544.81 | 39.304 | | | 156.163 | 12.497 | | |
| Delay (subjects) | | | | | 5.584 | 2.363 | -0.09 | |
| Sound (subjects) | 76.63 | 8.754 | 0.45 | | | | | |
| Residual | 3916.80 | 62.584 | | | 722.60 | 26.881 | | |

*Note*: $BF_{10}$: Bayes factor (values >1 indicate evidence for the alternative hypothesis; values <1 indicate evidence for the null hypothesis). Statistically significant *t*-values ($\geq 1.96$) and $BF_{10}$s meeting the 3/ 0.33 threshold (approx. corresponding to the 0.05 alpha level) are formatted in bold.